\documentclass[12pt]{iopart}
\usepackage{graphicx}

\begin{document}

\title{Reflection beamshifts of visible light due to graphene}

\author{Nathaniel Hermosa}

\address{National Institute of Physics, University of the Philippines, Diliman Quezon City 1101 Philippines}
\ead{nhermosa@nip.upd.edu.ph}

\begin{abstract}
I present calculations of reflection beamshifts, Goos-H\"anchen and Imbert-Fedorov shifts, due to the presence of a  monolayer graphene on a dielectric media when using a beam with wavelength in the visible range. Measuring the Goos-H\"anchen and Imbert-Fedorov shifts is an alternative method to determine graphene's conductivity. I look at beamshifts for different polarization states ($p$, $s$, $45^0$, $\sigma^+$) and  I discuss other possible experimental routes to determine these beamshifts and consequently, the graphene's optical conductivity. The Goos-H\"anchen shifts for visible light I calculated are in good agreement with results of a recent experiment.  
\end{abstract}

\maketitle

\section{Introduction}

A beam of light with finite waist experiences minute angular and spatial deviations from what is expected from the law of reflection when it strikes an index gradient. Known collectively as beamshifts, the Goos-H\"anchen (GH) and Imbert-Fedorov (IF) shifts are corrections to the lateral and the transverse position and skew of the reflection, respectively \cite{bliokh, aiello_role}.  These shifts may occur simultaneously or separately and their presence depends on the polarization of the incident beam \cite{aiello_role,LiUnified2006,GoosHanchen1947,Fedorov1955,Imbert1972}, the index gradient seen by the beam \cite{MeranoOpEx2007,AielloPRA2009,HermosaOL2011,MenardPRB2010,BermanPRE2002}, the divergence of the beam \cite{MeranoNatPhotonics2009} and on the modal structure of the beam \cite{BliokhOL2009,MeranoPRA2010,AilleoOl2011,DasguptaOptComm2006,GollaPramana2011,HermosaSPIE2011, HermosaOL2012}.  Bliok and Aiello offer an excellent review of beamshifts in Ref.\cite{bliokh}.

Measurement of beamshifts can offer an alternative method to determine optical properties of  materials.  It is nondestructive and is dynamic (i.e. it can instantenously change values with a change in the material property). There have been several experiments that reported beamshifts in different materials such as in metal \cite{AielloPRA2009, HermosaOL2011}, in semiconductors \cite{MenardPRB2010} and in materials with negative index of refraction \cite{BermanPRE2002}. 

Recently, Li et al.  measure a giant GH shift in total internal reflection in a graphene-dielectric interface \cite{li} . Their model does not completely describe the experimental results they obtained.  In this paper, the cause of this large GH shift is explained and some routes to determine the other beamshifts are presented. A measurement of these shifts can provide alternative ways of determining material properties of graphene.

Graphene, an atom thick sheet of carbon,  has been identified as an adaptable material in photonics and optioelectronic  applications due to its remarkable absorption and its highly flexible optical properties that can change drastically when it is electrically gated (see for example Ref.\cite{koppens}). The problem however, is that there is still a big debate on the optical properties of graphene, especially since the properties are sensitive to its immediate environment \cite{stauber}. Until now, there is no universal experimental accepted value of graphene's optical conductivity other than the predicted theoretical value for a clean graphene at half-filling at zero temperature ($\sigma=e^2/{4\hbar}$) where $e$ is the charge of the electron, and $\hbar$ is the Planck's constant \cite{luo, luo2}. This fact makes it imperative to study the optical conductivity of graphene. An alternative route that is dynamic and non-invasive will contribute significantly to this end. 

In this paper, I calculate reflection beamshifts for light sources in the visible range. I provide experimental insights on how to measure these shifts.
\section{Theoretical framework}
\subsection{Reflection beamshifts}
When a beam of light of finite transverse extent impinges on an index gradient, the beam may experience four beam shifts: two spatial shif ts ($\Delta_{GH}$ and $\Delta_{IF}$) and two angular shifts ($\Theta_{GH}$ and $\Theta_{IF}$). The derivation of these expressions is too long to be presented here and has been extensively discussed elsewhere \cite{bliokh}. However, it follows the general procedure of decomposition of the incident beam into plane wave components and the action of the Fresnel reflection coefficients to the s and p polarization components of these waves. Summing all the reflected plane wave components and taking the centroid of the intensity distribution gives the following dimensionless beamshifts: \begin{equation}
\Delta_{GH}=w_{p}\mathrm{Im}\left(\frac{\partial\ln r_{p}}{\partial\theta}\right)+w_{s}\mathrm{Im}\left(\frac{\partial\ln r_{s}}{\partial\theta}\right)
\end{equation}
\begin{equation}
-\Theta_{GH}=w_{p}\mathrm{Re}\left(\frac{\partial\ln r_{p}}{\partial\theta}\right)+w_{s}\mathrm{Re}\left(\frac{\partial\ln r_{s}}{\partial\theta}\right)
\label{eq:theta_GH}
\end{equation}
\begin{equation}
\Delta_{IF}=-\frac{a_{p}a_{s}\cot\theta}{R_{p}^{2}a_{p}^{2}+R_{s}^{2}a_{s}^{2}}\times\quad\left[\left(R_{p}^{2}+R_{s}^{2}\right)\sin\eta+2R_{p}R_{s}\sin\left(\eta-\varphi_{p}+\varphi_{s}\right)\right]
\end{equation}
\begin{equation}
\Theta_{IF}=\frac{a_{p}a_{s}\cot\theta}{R_{p}^{2}a_{p}^{2}+R_{s}^{2}a_{s}^{2}}\left[\left(R_{p}^{2}-R_{s}^{2}\right)\cos\eta\right],
\label{eq:theta_IF}
\end{equation}
\noindent{where $\Delta_{GH}$ and $\Theta_{GH}$ are the dimensionless spatial and angular GH shifts, respectively and $\Delta_{IF}$ and $\Theta_{IF}$ are the dimensionless spatial and angular IF shifts, respectively, with $w_{s/p}=R_{s/p}^{2}a_{s/p}^{2}/\left(R_{p}^{2}a_{p}^{2}+R_{s}^{2}a_{s}^{2}\right)$, and $r_{s/p}=R_{s/p}\exp\left(i\varphi_{s/p}\right)$, the Fresnel reflection coefficient evaluated at the incident angle $\theta$, the $\varphi_{s/p}$ is the phase gain after reflection, $a_{s/p}$ are the electric field components while $\eta$ is the relative phase difference between these components. The factors in the dimensionless beamshifts are independent of the wavelength $\lambda$ except for the materials' permittivity incorporated in the Fresnel coefficients. Since the beamshifts are dependent on the Fresnel coefficient, a change in it will change the beamshifts.}

The physical beamshifts $\Gamma_X$ and $\Gamma_Y$ are the sum of the contribution of the spatial $\Delta_{GH,IF}$ and angular $\Theta_{GH,IF}$ are given by,
\begin{equation}
k_{0}\Gamma_X=\Delta_{GH}+(z/L)\Theta_{GH}
\end{equation}
\begin{equation}
k_{0}\Gamma
_Y=\Delta_{IF}+(z/L)\Theta_{IF},\end{equation}

\noindent{respectively, where $k_{0}=2\pi/\lambda$, $z$ is the distance of the detector from the minimum beam waist and $L=k_0\omega^2_0/2$ is the Rayleigh length.} 

One can use the scheme developed by Woerdman et al., to detect these shifts  (see for example \cite{AielloPRA2009,HermosaOL2011,MeranoNatPhotonics2009}).  In that method, the polarization differential beamshift is measured with a quadrant detector while toggling between two orthogonal polarizations and reducing technical noise with a lockin amplifier. The difference in the intensity of the signal as detected by the quadrant detector gives the value of the differential beamshifts between the two polarization states.  

On the other hand, weak measurements can also be used in determining beam shifts (see for example \cite{luo,luo2,luo3}).   As a matter of fact, weak measurement can amplify these shifts via weak amplification. However, the amplification factor has been found to be not constant \cite{luo2}, hence an \textit{apriori} knowledge is necessary and may not be as useful when material properties are unknown.
 
\subsection{Fresnel coefficients of a monolayer graphene}

The Fresnel coefficient for a dielectric-graphene-dielectric interface can be solved by imposing the following boundary conditions, \textbf{n}$\times(E_{i+1}-E_{i})|_{z=0}=0$ and \textbf{n}$\times(H_{i+1}-H_{i})|_{z=0}=J$ where \textbf{n} is the unit surface normal, $E_{i,i+1}$ and $H_{i,i+1}$ are the Electric fields and Magnetic fields at interface and $J$ is the surface current density of the graphene that is proportional to its conductivity $ \tilde{\sigma}$\cite{jackson}. The resulting reflection coefficients are

\begin{equation}
r_s=\frac{\sqrt{\epsilon_1}\cos\theta_1-\sqrt{\epsilon_2}\cos\theta_2- \tilde{\sigma}/{\epsilon_{0}c}}{\sqrt{\epsilon_1}\cos\theta_1+\sqrt{\epsilon_2}\cos\theta_2+ \tilde{\sigma}/{\epsilon_{0}c}}
\end{equation}
\begin{equation}
r_p=\frac{\sqrt{\epsilon_2}/\cos\theta_2-\sqrt{\epsilon_1}/\cos\theta_1+ \tilde{\sigma}/{\epsilon_{0}c}}{\sqrt{\epsilon_2}/\cos\theta_2+\sqrt{\epsilon_1}/\cos\theta_1+ \tilde{\sigma}/{\epsilon_{0}c}},
\end{equation}
\noindent{where $\epsilon_{1,2}$ are the dielectric constant of the material above and below the graphene, $\epsilon_0$ is the permittivity of free space and $c$ is the speed of light.  The $r_{s,p}$ equations are derived in \cite{koppens,zhan},  independently.}

The conductivity of graphene I use to calculate the GH and IF shifts in the visible range is $\tilde{\sigma}=e^2/4\hbar$ even though the conductivity of a real graphene has dependence on the temperature and the wavelength of the incident light, and on the Fermi level or the amount of doping.  This value, however, is not implausible experimentally as Kuzmenko et al measured a conductivity that is close to the predicted conductivity of a clean graphene \cite{kuzmenko} and that at high photon energies the conductivity of a clean graphene approaches this value even at finite temperatures \cite{peres}. Moreover, Mak observed that at visible light photon energies, the conductivity approaces this value even with scattering rate of 20eV and chemical potential of 100eV \cite{mak}. For calculations of the beamshifts in the visible range, I assume a graphene on a BK7 substrante ($n^2_{BK7}= \epsilon_{BK7}=2.295$) with air ($n^2_{air}= \epsilon_{air}=1$) on the other surface. 

\section{Results and Discussion}

\begin{figure}[h!]
\centering
 {\includegraphics[width=0.8\linewidth]{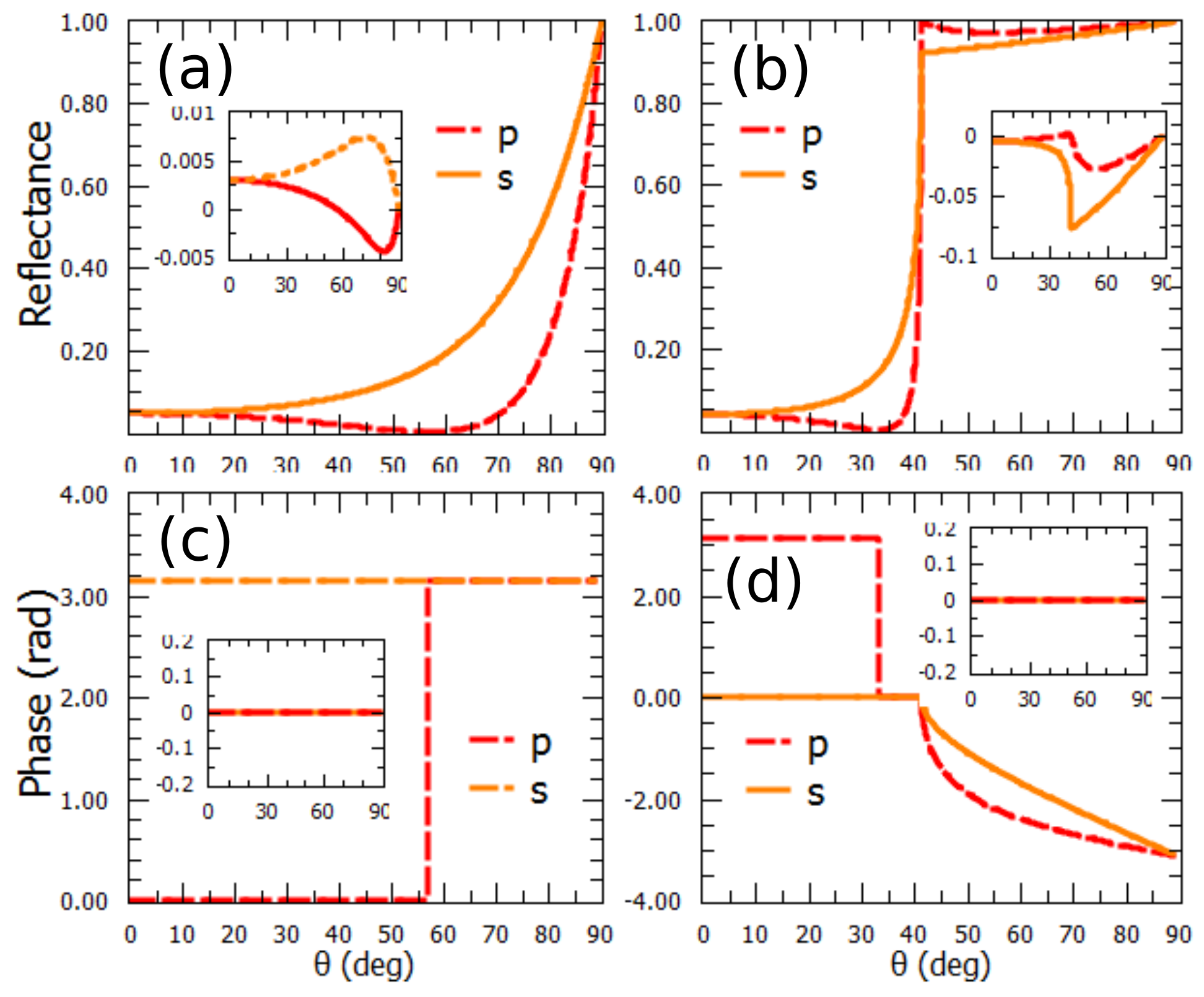}}
 \caption{(a) The reflectance during external reflection and (c) the corresponding phase jump in the reflected beam. (c) The reflectance when light comes from the dielectric and (d) the value of the phase jump of the reflected beam. Insets show the difference in the reflectance of the interface with and without the graphene.}
\label{1}
\end{figure}

Figure \ref{1} shows the reflectance and the amount of phase jumps after the incident beam is reflected at an angle $\theta$ on an air-graphene-glass/ glass-graphene-air interface.  When light strikes from air (Fig. \ref{1} (a) and (c)), the reflectance is very similar to the no graphene sample. There is a minute difference in the $R_{s,p}^2$ while it acquires the same amount of phase jumps as in the no graphene interface. These reflectances and phase jumps indicate that there are differences between the beamshifts for a graphene on top of a glass substrate compared to the bare substrate.  Also, the Brewster's angle has been shifted by $\sim$ $0.6^0$ due to the presence of the graphene. This is particularly helpful when measuring $\Theta_{GH}$ as it increases rapidly near the Brewster's angle.

The $R_s^2$ and $R_p^2$ with the graphene do not immediately reach the value of 1 after the critical angle (Fig. \ref{1}(b)) compared to the no graphene sample where  $R_{s,p}^2=1$ beyond and at the critical angle. There is a slow sloping $R_{s,p}^2$. This fact is important since angular GH may be present after the critical angle for a glass with graphene. The amount of phase jump after reflection however, are almost similar (Fig.\ref{1}(d)). 

\begin{figure}[h!]
\centering
 {\includegraphics[width=0.8\linewidth]{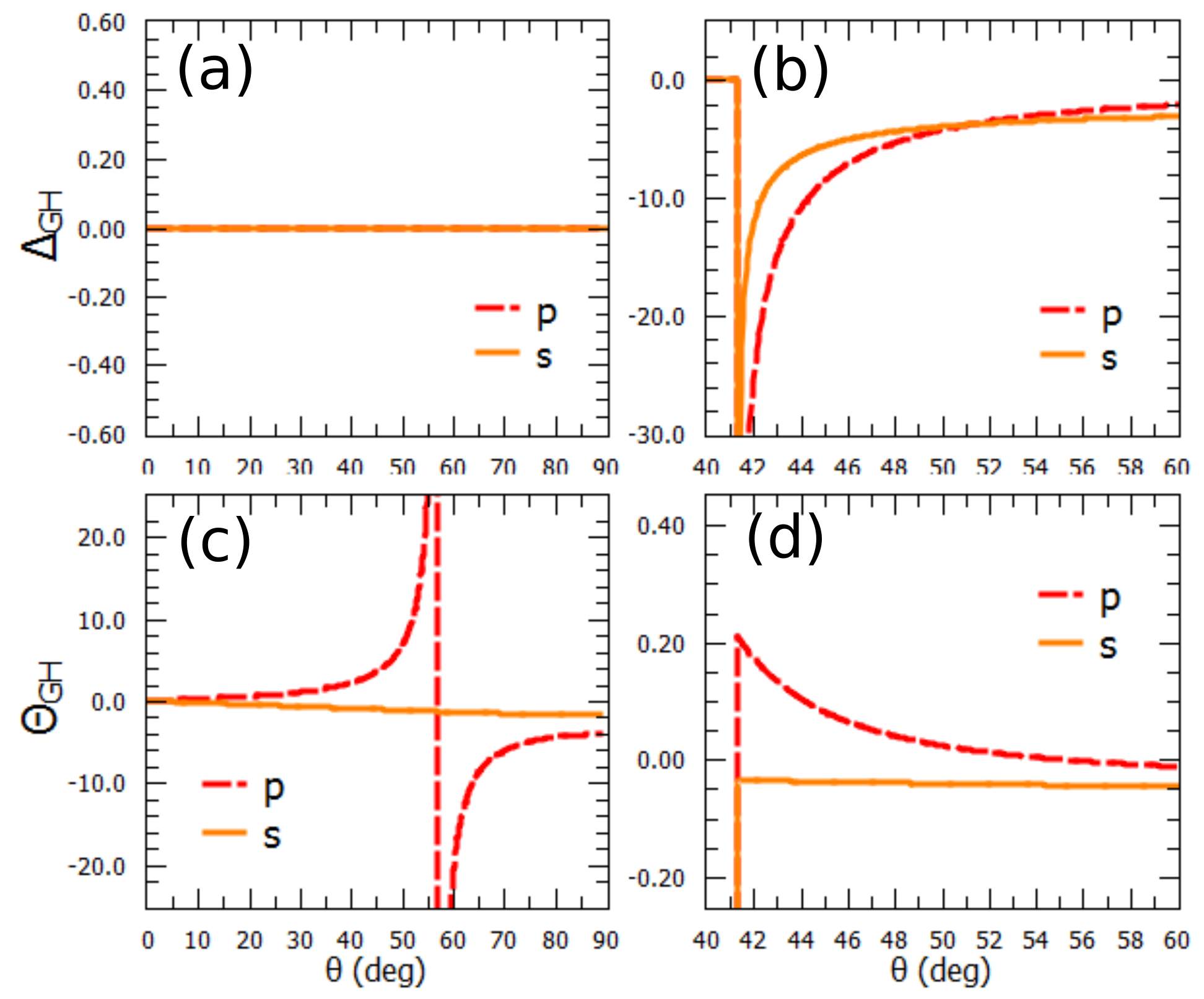}}
 \caption{Dimensionless GH shifts for polarization states $p$ and $s$. Spatial shifts $\Delta_{GH}$ for (a) external and  (b) internal reflection and angular shifts $\Theta_{GH}$ for (c) external and (d) internal reflection. (\emph{See text for detail.})}
\label{2}
\end{figure}

Figure \ref{2} shows the dimensionless spatial GH shifts ((a) and (b)) and angular GH shifts ((c) and (d)).  The main results here are the observance of a nonzero  $\Theta_{GH}$ for $p$ and $s$  linear polarization polarization states (fig.\ref{2}(d)) when light strikes the graphene film from the glass side.  These do not occur for a bare dielectric.  This fact which has not been reported in any literature before, means that by focusing the beam and letting it propagate further, the beamshifts can be amplified \cite{aiello_role}.  

In \cite{li}, they have observed a giant GH shift which decays slower with a focused beam compared to the nonfocused beam case. In their experiments, they measured the GH shifts for focused beam as position differential between the shift of $p$ and $s$ with and without graphene, with two balanced amplified photodetectors. Their beam is a HeNe laser (wavelength $\lambda$ = 632.8 $nm$) with  an  initial waist of 1 $mm$ that is focused by a lens with a 2 $mm$ focal length. The researchers needed to focus the beam to achieve high measurement accuracy. Aiello et al., describing the role of a lens in the measurement of the shifts in ref. \cite{aiello_role}, note that the lens changes the nature of the shift, either it becomes a skew or a spatial deflection or a combination depending on the placement of the lens. The general behavior of the shift however, with respect to the incident angle will not be affected. Although my calculations did not consider a lens after reflection, I obtained a similar trend: a gradual roff-off behavior with respect to the incident angle. Since the experimental parameters in \cite{li} are not complete, a thorough comparison cannot be fully made. The order of magnitude and the behavior strongly indicate that the cause of the experimental results is indeed  $\Theta_{GH}$.

 Moreover, I calculated small spatial shifts $\Delta_{GH}$ at polarization states $p$ and $s$, (Fig. \ref{2}(c)) that follow the Artmann formula \cite{artmann}. The values however, are very close to that of the bare sample to be distinguishable in experiments (e.i. $\Gamma_X$ $\sim$ fraction of a nanometer). 

In the vicinity of the Brester's angle, the  $\Theta_{GH}$ magnitude is huge (Fig. \ref{2}(c)) when light strikes from air. This can be exploited when determining graphene's conductivity using beamshift with a $p$-polarized beam. However at very near the Brewster's angle, the intensity will be diminished which will make it difficult to detect, there is parasitic signal from cross-polarization \cite{aielloOL2009} and it will be troublesome as the theoretical equation derived in Ref. \cite{bliokh} is not sufficient. The $\Theta_{GH}$ when using $s$- polarized light also give easily detectable beamshifts (Fig. \ref{2}(c)). 

 \begin{figure}[h!]
\centering
 {\includegraphics[width=0.8\linewidth]{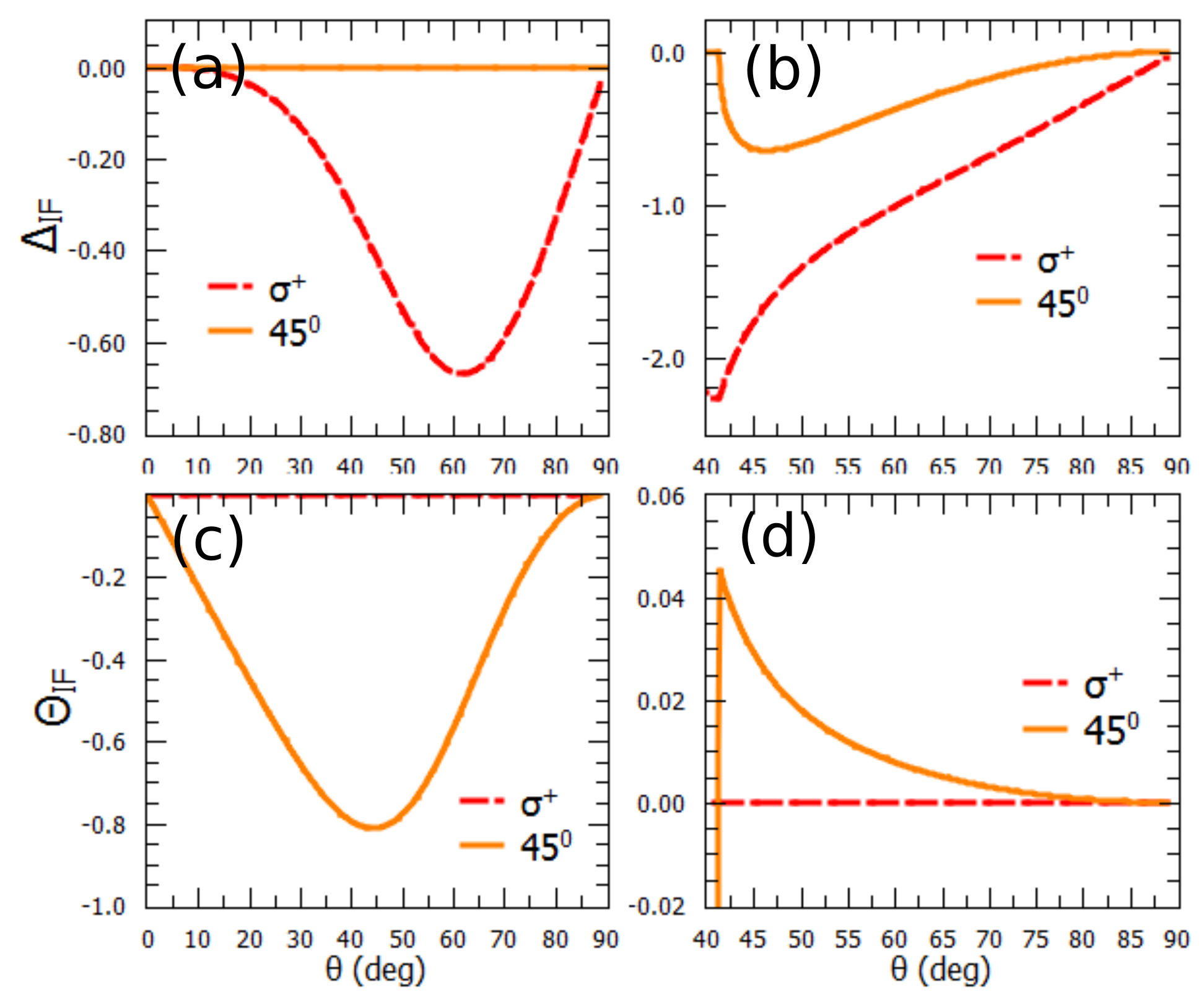}}
 \caption{Dimensionless IF shifts for polarization states $\sigma^+$ and $45^0$. Spatial shifts $\Delta_{IF}$ for (a) external and  (b) internal reflection and angular shifts $\Theta_{IF}$ for (c) external and (d) internal reflection. (\emph{See text for detail.})}
\label{3}
\end{figure}

The dimesionless spatial IF shifts ((a) and (b)) and angular IF shifts (c) are expected for all polarization state, again, except for the angular IF shifts for $45^0$ polarization (d). The $\Delta_{IF}$ happens only for $\sigma^+$(fig. \ref{2}(a)) and the $\Theta_{IF}$ happens only for $45^0$(fig. \ref{2}(c)) for external reflection. Measuring $\Delta_{IF}$ could be daunting because at maximum difference, it is only about 2.5 nm more that in the case of $\Delta_{IF}$ with a bare substrate. The calculation of non-zero $\Theta_{IF}$ similar to $\Theta_{GH}$ in internal reflection have not been reported in literature. This could be exploited in determining the optical conductivity of graphene.

It will be instructive to give the differential beamshifts in physical units with and without the graphene ($\Delta\Gamma_{X,Y}=\Gamma_{X,Y}^{graphene}-\Gamma_{X,Y}^{bare}$). Here, I calculated the shifts using a beam with HeNe laser wavelength ($\lambda$ = 632.8nm), a detection distance of 23 cm from the focus of a lens ($f$=70mm) and a beam waist of $\omega_0$=20$\mu$m. The beamshifts will add up as the $\Delta_{GH}$ ($\Delta_{IF}$) and $\Theta_{GH}$ ($\Theta_{IF}$) both manifest as a minute movement of the beam. The difference in the beamshifts due to the $\Delta$'s and $\Theta$'s is that the latter grows linearly with propagation. For internal reflection,  the maximum differential shift in the longitudinal direction $\Delta\Gamma_X^{max}$ occurs in p-polarized light and is in the order of  micrometers (as measured by \cite{li}), while the maximum differential shift in transverse direction $\Delta\Gamma_Y^{max}$ occurs in $45^0$ polarized light in the order of hundreds of nanometers. In the case of external reflection, $\Delta\Gamma_X^{max}$ happens at p-polarized beam (also in the order of micrometers). In all these beamshifts, the $\Theta$ shifts dominate the $\Delta$ shifts. These can easily be detected with a quadrant cell.

The $\Theta_{GH}$ and the $\Theta_{IF}$ during internal reflection can be exploited to measure the $\tilde\sigma$. Near the critical angle, I have calculated the expression for $\Theta_{GH}$ from eqn. \ref{eq:theta_GH} as,

\begin{equation}
\Theta_{GH}\approx \frac{\tilde\sigma}{n\epsilon_0c}\frac{2\cos^3(\theta)\sin(2\theta)}{\left(\left(n^2\sin^2(\theta)-1\right)+\left(\frac{\cos^2(\theta)}{n}\right)^2\right)^2},
\label{eq:approx_GH}
\end{equation}

\noindent where $n$ is the index of refraction of the substrate. In this expression, $\tilde\sigma$ is just a constant factor which can be use as a parameter. Equation \ref{eq:approx_GH} gives values within $10\%$ of the values without the approximation, within $\sim 8^0$ after the critical angle. As an order of magnitude comparison, the physical beam shift due to $\Theta_{GH}$ given the parameters above is 1 to 2 orders of magnitude greater than the correction due to the approximation.

The simplified expression for $\Theta_{IF}$ from eqn. \ref{eq:theta_IF} is given by,

\begin{equation}
\Theta_{IF}\approx\frac{\tilde\sigma}{\epsilon_0c}\frac{n\sin\left(2\theta\right)\sin\theta\left(n^2-1\right)}{\left(n^6-n^4-n^2+1\right)\sin^2\theta-n^4+2n^2-1}.
\label{eq:approx_IF}
\end{equation}

Again, the $\tilde\sigma$ is just a factor in $\Theta_{IF}$. Eqn. \ref{eq:approx_IF} is within 5\% of the value with the approximation for all angles greater than the critical angle. With the experimental parameters given above, the correction to the physical beam shift due to the approximated $\Theta_{IF}$ given in eqn.\ref{eq:approx_IF} is in the order of a few nanometers.  

The values of the physical beamshifts due to $\Theta_{GH}$ and $\Theta_{IF}$ are within experimental resolution even without the need to use weak amplification. The values in \cite{HermosaOL2011,MeranoNatPhotonics2009, li} have similar order of magnitudes in my calculations here.

\section{Conclusion}
\label{sec:conclusion}

In this paper, I have presented calculations for reflection beamshifts, Goos-H\"anchen and Imbert-Fedorov shifts, for visible light when a monolayer of graphene is placed on a dielectric surface. The four beamshifts can be present depending on the polarization of the incident beam. The spatial Goos- H\"anchen and Imbert-Fedorov shifts, $\Delta_{GH}$ and $\Delta_{IF}$ for intenal reflection, as well as the angular  Goos- H\"anchen  and Imbert-Fedorov shifts, $\Theta_{GH}$ and $\Theta_{IF}$ shifts for external reflection are quite similar in behavior to their counterpart shifts with bare substrate. The main results here are the nonzero $\Theta_{GH}$ and $\Theta_{IF}$ even at angles beyond the critical angle under internal reflection. I calculated shifts that can be measured relatively easily in experiments. I also derived expressions for the $\Theta_{GH}$ and $\Theta_{IF}$ shifts that isolate the optical conductivity $\tilde\sigma$ from known factors such as the incident angle, $\theta$ and the index of refraction of the substrate. These approximations are well within 10\% of the non approximated value.  Measuring beam shifts could be an alternative way to determine the optical conductivity of graphene.

\section*{Acknowledgement}
\label{sec:acknowledgement}

N. Hermosa is a Universitiy of the Philippines Office of the Vice President for Academic Affairs Balik PhD program recipient (UP OVPAA 2015-06).

\end{document}